\documentclass[aps,prd,amsmath,amssymb,twocolumn,preprintnumbers,nofootinbib]{revtex4-1}

\usepackage{graphicx}
\usepackage{dcolumn}
\usepackage{bm}
\usepackage[bookmarks, pagebackref=false]{hyperref}
\usepackage[usenames,dvipsnames]{xcolor}
\definecolor{rossoCP3}{cmyk}{0,.88,.77,.40}
\definecolor{blaa}{rgb}{0.2,0.2,0.6}
\hypersetup{colorlinks, 
	bookmarksopen, 
	bookmarksnumbered,
	citecolor=blaa, 		
	linkcolor=rossoCP3,	
	urlcolor=rossoCP3,			
}

\usepackage{amsmath}
\usepackage{slashed}
\usepackage{tikz}
\usepackage{graphicx}
\usepackage{pgfplots}
\usepackage{subfigure}
\usepackage{color}
\usepackage{rotating}
\usepackage{gensymb}
\usepackage{simplewick}
\usepackage{enumerate}
\usepackage{soul,xcolor}

\usetikzlibrary{quotes,angles,snakes}
\usetikzlibrary{trees}
\usetikzlibrary{decorations.pathmorphing}
\usetikzlibrary{decorations.markings}
\usetikzlibrary{arrows,shapes,positioning}
\usetikzlibrary{decorations.markings}
\tikzstyle arrowstyle=[scale=1]
\tikzstyle directed=[postaction={decorate,decoration={markings,
    mark=at position .65 with {\arrow[arrowstyle]{stealth}}}}]
\tikzstyle reverse directed=[postaction={decorate,decoration={markings,
    mark=at position .65 with {\arrowreversed[arrowstyle]{stealth};}}}]
\DeclareMathOperator{\Tr}{Tr}
\newcommand{\RNum}[1]{\uppercase\expandafter{\romannumeral #1\relax}}

\newcommand{\beq}{\begin{equation}}
\newcommand{\eeq}{\end{equation}}
\newcommand{\bea}{\begin{eqnarray}}
\newcommand{\eea}{\end{eqnarray}}
\usepackage{ulem}

\newcommand\vRto{\mathrel{\stackrel{\makebox[0pt]{\mbox{\normalfont\tiny \text{$v_R$}}}}{\text{$\longrightarrow$}}}}

\newcommand{\trace}[1]{\ensuremath{\mathrm{Tr}#1}}

\usepackage{xcolor,colortbl}
\definecolor{Blue}{RGB}{140,165,195}
\definecolor{Purple}{RGB}{255,145,145}
\definecolor{bluc}{cmyk}{1,1,0,0.1}
\definecolor{rossoCP3}{cmyk}{0,.88,.77,.40}
\definecolor{rosso}{cmyk}{0,1,1,0.4}
\definecolor{rossos}{cmyk}{0,1,1,0.55}
\definecolor{rossoc}{cmyk}{0,1,1,0.2}
\definecolor{verdes}{cmyk}{0.92,0,0.59,0.4}

\usepackage{color}
\usepackage{ulem}

\begin{document}

\setstcolor{red}

\newcommand{\blue}[1]{{\color{blue}#1}}
\newcommand{\red}[1]{{\color{red}#1}}
 
\title{ \LARGE  \color{rossoCP3} Gravitational Waves from Pati-Salam  Dynamics}
\author{W.C.~Huang$^{\color{rossoCP3}{\heartsuit}}$}
\author{F.~Sannino$^{{\color{rossoCP3}{\heartsuit}},{\color{rossoCP3}{\spadesuit}}}$
}
\author{Z.W.~Wang$^{\color{rossoCP3}{\heartsuit}}$}

\affiliation{$^{\color{rossoCP3}{\heartsuit}}$ {\mbox {$\rm{CP}^3$-Origins, University of Southern Denmark, Campusvej 55}}
5230 Odense M, Denmark \\
}
\affiliation{\mbox{ $^{\color{rossoCP3}{\spadesuit}}$Dipartimento di Fisica “E. Pancini”, Università di Napoli Federico II | INFN sezione di Napoli}\\ \mbox{Complesso Universitario di Monte S. Angelo Edificio 6, via Cintia, 80126 Napoli, Italy.}}

\begin{abstract}  

We show that it is possible to use gravitational wave detectors to observe the occurrence of a  first order phase transition in Pati-Salam extensions of the Standard Model.   We show that the peak frequency of the expected gravitational wave signals ranges within $0.1-10$ Hz.  We find amusing that the next generation of gravity waves detectors are able to explore time honoured extensions of the Standard Model occurring at energy scales inaccessible by present and future particle physics accelerators. 
\\
[.3cm]
{\footnotesize  \it Preprint: CP$^3$-Origins-2020-05 DNRF90}
\end{abstract}
\maketitle

\section{Introduction}

The idea of using gravitational wave as a complementary approach to explore particle physics started some time ago e.g.~\cite{Apreda:2001us,Grojean:2006bp,Jarvinen:2009mh}. However, the bulk of the research concentrated, so far, on the electroweak phase transition which is typically in the detection region of the LISA gravitational wave detector  as nicely summarised in \cite{Caprini:2015zlo,Caprini:2019egz}. Of special interest is the possible detection of gravity waves originated in Grand Unified Theories (GUT). The interest arises also because typically the new physics energy scale of GUTs is beyond the reach of the existing and even future $100\,\rm{TeV}$ colliders. 

A prerequisite to start even discussing gravitational wave detection is that the underlying theory must undergo a strong first order phase transition at some point during the evolution of the Universe. Additionally the higher is the energy scale of the first order phase transition the higher will be the peak frequency of the gravitational wave that needs to be detected. Inevitably, the upper frequency limit of the existing and planned gravitational wave detectors (roughly at order of $10^3\,\rm{Hz}$) provides an upper bound on the detectable energy scale (roughly at $10^4\--10^5\,\rm{TeV}$). In this sense, among the different types of GUTs, only the two semi-simple GUTs: Pati-Salam model \cite{Pati:1974yy} and Trinification model \cite{Babu:1985gi} satisfy this criterion. In this work, we will focus mainly on the gravitational wave signatures of the minimal Pati-Salam model. Our investigation differs from the one in  \cite{Croon:2018kqn} in which an alternative model of Pati-Salam model was considered. In that work the authors employed a rather involved matter content that featured, however, a simpler first order phase transition structure \footnote{In the work of \cite{Croon:2018kqn}, the authors try to realize the gauge coupling unification and symmetry breaking to an intermediate step (left-right model) first and thus their scalar sectors are overall more complicated. However, their first order phase transition occurs only when $SU(4)$ is breaking while in our case both $SU(4)$ and $SU(2)_R$ breaks and thus their analysis of the first order phase transition is simpler and fewer couplings are involved.}. 

The Pati-Salam model of matter field unification \cite{Pati:1974yy} is a time-honoured example in which one can address the hypercharge triviality issue by embedding it in an asymptotically free theory. From a phenomenological standpoint it can be commended because it does not induce fast proton decay, and it can even be extended to provide a stable proton \cite{FileviezPerez:2016laj} while automatically providing a rationale for the existence of right handed neutrinos (see more details in a recent nice review \cite{Pati:2017ysg}). 

So far, asymptotic freedom has been the well traveled route to resolve the triviality problem. An alternative route is that in which the UV theory acquires an interacting fixed point, before gravity sets in, de facto {\it saving} itself from the presence of a cutoff.  This unexplored route was opened when the first safe gauge-Yukawa theory was discovered in \cite{Litim:2014uca}. 

To achieve a safe theory with a small number of colours we employ large number of matter fields techniques \cite{PalanquesMestre:1983zy,Gracey:1996he}. The first phenomenological applications of the large $N_f$ limit appeared in \cite{Mann:2017wzh} where it was first explored whether the SM augmented by a large number of vector-like fermions can have an ultra-violet fixed-point in all couplings. The full treatment appeared in \cite{Pelaggi:2017abg} and further generalized in \cite{Antipin:2018zdg}.
It was found in \cite{Pelaggi:2017abg} and later on proved in \cite{Antipin:2018zdg} that while the non-abelian gauge couplings, Higgs quartic and Yukawa coupling can exhibit a safe fixed point, the hypercharge remains troublesome. In fact, for abelian  theories  the fermion mass anomalous dimension diverges at the alleged fixed point \cite{Antipin:2017ebo} suggesting that a safe extension of the SM, like the asymptotically free counterpart, is best obtained by embedding the SM in a non-abelian gauge structure.
The first non-abelian safe PS and Trinification embeddings were put forward in  \cite{Molinaro:2018kjz, Wang:2018yer}.  However, in the minimal models, only one generation of SM fermions can be modelled, since all the Yukawa couplings are determined by the same UV fixed point value with no resulting hierarchy at low energy.  Yukawa hierarchies among three generations of SM fermions are discussed in \cite{Sannino:2019sch}.

In this work, we will start by  investigating  gravitational wave signatures emerging in   Pati-Salam extensions of the Standard Model  embedded in an asymptotically safe scenario. We use these predictions as an initial seed value to study the first order phase transition and gravitational wave signatures. Later, we will depart from the safety scenario and will  explore a more general parameter space. 
Therefore, our work of studying the phase transition and gravitational wave generation is very general and applies to
both safe and non-safe embeddings of the Pati-Salam model.

We discover that the next-generation gravity waves detectors are  able to explore time honoured extensions of the Standard Model occurring at energy scales inaccessible by present and future particle physics colliders. More precisely we show that the peak frequency of the expected gravitational wave signals ranges within $0.1-10$ Hz.

The paper is organised as follows. In Section~\ref{PS-model} we introduce the Pati-Salam model while in Section~\ref{Temperature} we compute the finite temperature corrections to the relevant part of the potential of the theory. The order of the phase transition as well as gravitational waves generation and detection are studied in Section~\ref{PhaseTransition}. The predictions for the gravity waves signals stemming from the model parameters are presented in Section~\ref{GravityWaves}. We conclude in Section~\ref{Conclusions}. In the appendix we provide some detailed computations. 

\section{Introducing the Pati-Salam model}
\label{PS-model}

We first briefly review the Pati-Salam embedding of the SM suggested in \cite{Molinaro:2018kjz}.

Consider the time-honored PS gauge symmetry group $G_\text{PS}$ \cite{Pati:1974yy}
\begin{equation}
	G_\text{PS} = SU(4)\otimes SU(2)_{L} \otimes SU(2)_{R}\,,\label{Pati_Group}
\end{equation}
with gauge couplings $g_4$, $g_L$ and $g_R$, respectively. Here the gauge group $SU(4)\supset SU(3)_C \otimes U(1)_{B-L}$, where $SU(3)_C$ denotes the SM QCD gauge group.
The SM quark and lepton fields are unified into the $G_{\rm PS}$ irreducible representations
\begin{eqnarray}
\begin{split}\label{fermionsLR}
	\psi_{L i} &= \left(\begin{array}{cccc} u_L^1  & u_L^2 & u_L^3 & \nu_L\\ d_L^1  & d_L^2 & d_L^3 & e_L\end{array}\right)_i \sim  (4,2,1)_i \,, \\ 
	\psi_{R i} &= \left(\begin{array}{cccc} u_R^1  & u_R^2 & u_R^3 & \nu_R\\ d_R^1  & d_R^2 & d_R^3 & e_R\end{array}\right)_i \sim (4,1,2)_i \,,  
\end{split}
\end{eqnarray}
where $i=1,2,3$ is a flavor index. 
In order to induce the breaking of $G_{\rm PS}$ to the SM gauge group, we introduce a scalar field $\phi_R$ which transforms as the fermion multiplet $\psi_R$, that is, $\phi_R\sim (4,1,2)$: 
\begin{equation}
	\phi_R \; = \; \left(\begin{array}{cccc}  \phi_R^{u_1} & \phi_R^{u_2} &\phi_R^{u_3} & \phi_R^0 \\ \phi_R^{d_1}  & \phi_R^{d_2}  &\phi_R^{d_3}  & \phi_R^- \end{array} \right)\,,
\end{equation}
where the neutral component $\phi_R^0$ takes a non-zero vev, $\langle\phi_R^0 \rangle \equiv v_R$, such that $G_{\rm PS} \vRto SU(3)_C \otimes SU(2)_L \otimes U(1)_Y$.
We also introduce an additional (complex) scalar field $\Phi\sim (1,2,2)$, with
\begin{eqnarray}
	\Phi & = & \left(\begin{array}{cc} \phi_1^0 & \phi_2^+ \\ \phi_1^- & \phi_2^0 \end{array} \right) \equiv \left(\begin{array}{cc} \Phi_1 & \Phi_2 \end{array}\right)\,,
\end{eqnarray}
which is responsible of the breaking of the EW symmetry.

The  most general Yukawa Lagrangian for the matter fields $\psi_{L/R}$ is:
\begin{eqnarray}
\mathcal{L}_{\rm Yuk}^\psi & = & y \, \text{Tr}\left[\overline{\psi_L}\, \Phi\, \psi_R \right] \,+\, y_c \, \text{Tr}\left[\overline{\psi_L}\, \Phi^c\, \psi_R \right]\,+\, \text{h.c.}\,,\label{LYuk1}
\end{eqnarray}
where $y$ and $y_c$ are the Yukawa couplings for the third generation only. Note that the Yukawa couplings of the first two generations can be generated through the clockwork mechanism \cite{Sannino:2019sch}.

In the case of a self-conjugate bi-doublet field $\Phi \equiv \Phi^c$,  one obtains degenerate masses at tree-level, namely
\begin{equation}
	m_t = m_b = m_\tau = m_{\nu_\tau}\,.\label{spectrum2}
\end{equation}
 In order to separate the neutrino and top masses in Eq.~(\ref{spectrum2}), we implement the seesaw mechanism 
\cite{Minkowski:1977sc,Yanagida:1979as,GellMann:1980vs,Mohapatra:1979ia} by adding a new chiral fermion singlet  $N_L\sim (1,1,1)$, which has Yukawa interaction (see e.g.~\cite{Volkas:1995yn,Molinaro:2018kjz} for more details)
\begin{equation}
	\mathcal{L}_{\rm Yuk}^N \; = \; - y_\nu \,\overline{N_L} {\rm Tr}\left[ \phi_R^\dagger \,\psi_R \right]\,+\,{\rm h.c.}
\end{equation}
In order to split the mass of top, bottom and tau lepton in Eq.~\eqref{spectrum2}, we introduce a new vector-like fermion $F\sim (10, 1, 1)$ with mass $M_F$ and Yukawa interactions (see e.g.~\cite{Volkas:1995yn,Molinaro:2018kjz} for more details):
\begin{eqnarray}\label{YukF}
	\mathcal{L}_{\rm Yuk}^F & = & y_F\, {\rm Tr}\left( \overline{F_L} \, \phi_R^T \,i \tau_2 \,\psi_R \right) \,+\, {\rm h.c.}	\label{Yuk_F}
\end{eqnarray}
All the field contents and couplings are summarized in Tab.~\ref{couplings}.
\begin{table}[t!]
\centering
  \begin{tabular}{|| l | l | l ||}
    	\hline
Gauge & Yukawa & Scalar \\ \hline
$SU(4):\,g_4$ & $\psi_{L/R}: \,y,\,y_c$ & $\phi_R:\,\lambda_{R1},\,\lambda_{R2}$\\\hline
$SU(2)_L:\,g_L$ & $N_L:\,y_{\nu}$ & portal: $\lambda_{R\Phi_1},\,\lambda_{R\Phi_2},\,\lambda_{R\Phi_3}$\\ \hline
$SU(2)_R:\,g_R$ & $F :\,y_F$ & $\Phi:\,\lambda_1,\,\lambda_2,\,\lambda_3,\,\lambda_4$\\ \hline
\end{tabular}
\caption{\small Gauge, Yukawa and scalar quartic couplings of the PS model.}
\label{couplings}
\end{table}

\section{Finite Temperature Effective Potential}
\label{Temperature}
\subsection{Tree Level Effective Potential of Pati-Salam Model}
The relevant terms in the tree level effective potential can be written as:
\begin{equation}
V_{\rm{tree}}\left(\phi_R\right)=\lambda_{R1} \trace^2\left(\phi_R^\dagger\phi_R \right)
				+ \lambda_{R2}\,\trace\left( \phi_R^\dagger\phi_R \phi_R^\dagger\phi_R \right) \,.\label{potential tree}
\end{equation}
It is important to note that we do not include any explicit mass terms in the tree level potential. The symmetry breaking in this work is induced by Coleman-Weinberg mechanism.

If we write out $\phi_R$ explictly as :
\begin{equation}
\frac{1}{\sqrt2}\left(
\begin{array}{cccc}
 \phi_{R1}+i\phi_{R2} & \cdots &\cdots & \phi_{R7}+i\phi_{R8} \\
 \phi_{R9}+i\phi_{R10}& \cdots &\cdots & v+\phi_{R15}+i\phi_{R16} \\
\end{array}
\right) \, ,
\end{equation}
where we choose the symmetry breaking direction of $\phi_R$ and thus all field components except the $\phi_{R15}$ direction are   zero.  As mentioned above, $\langle \phi_R \rangle $ triggers breaking of $G_{\rm PS} \vRto SU(3)_C \otimes SU(2)_L \otimes U(1)_Y$.
Out of sixteen scalar fields, there are nine Goldstone bosons and seven physical bosons.
Therefore, eight gauge bosons of $SU(4)$ (corresponding to QCD gluons) and one gauge field from $SU(4) \otimes SU(2)_R$
(which is simply $U(1)_Y$, a linear combination of the $U(1)_{B-L}$ from $SU(4)$ and $U(1)_R$ from $SU(2)_R$, with $Y=2I_R+B-L$) remain massless. The other nine gauge bosons of $SU(4) \otimes SU(2)_R$ (six lepto-quark, two right boson $W_R^{\pm}$ and one $Z'$) become massive.

With Eq.\eqref{potential tree}, we can construct the mass matrix of the scalar fields and obtain sixteen tree level mass eigenvalues.
These mass eigenvalues can be divided into nine Goldstone bosons with a mass $M_{\rm{Gold}}^2=v^2 \left(\lambda _{\text{R1}}+\lambda _{\text{R2}}\right)$ and seven physical Higgses, one out of which has a mass of $M_{\rm{Higgs1}}^2=3 v^2 \left(\lambda _{\text{R1}}+\lambda _{\text{R2}}\right)$ and six other Higgses with a  mass $M_{\rm{Higgs2}}^2=v^2 \lambda _{\text{R1}}$.

\subsection{Loop Level Effective Potential of Pati-Salam Model}
In this section, we will discuss the one loop contributions to the effective potential from scalar, gauge fields and fermions. The general formula is well known and can be written as:
\begin{equation}
V_{\rm{1loop}}=\sum_{i}\pm n_i\frac{m_i^4}{64\pi^2}\left(\log\left[\frac{m_i^2}{\mu^2}\right]-C_i\right)
\end{equation}
where the sum runs over the bosons $\left(+\right)$ and fermions $\left(-\right)$ and $n_i$ counts the internal degrees of freedom (d.o.f.) of each species $i$.
The symbols $m_i$, $\mu$ and $C_i$ correspond respectively to the tree level mass terms, renormalization scale and constant (equal to $5/6$ for gauge bosons and $3/2$ for scalars and fermions in Minimal Subtraction Scheme). 
We define the background field as $\rho$. In the following, we write out the scalars, gauge fields and fermions contribution explicitly.

The Higgs fields contributions (7 d.o.f.) to the one loop effective potential $V_{\text{Higgs}}$ are:
\begin{equation}
\begin{split}
&\frac{1}{64 \pi ^2}\left(3 \rho ^2 \left(\lambda _{R1}+\lambda _{R2}\right)\right)^2 \left(\log \left(\frac{3 \rho ^2 \left( \lambda _{R1}+\lambda _{R2}\right) }{\mu ^2}\right)-\frac{3}{2}\right)\\
&+\frac{6}{64 \pi ^2} \left(\rho ^2 \lambda _{R1}\right){}^2 \left(\log \left(\frac{\rho ^2 \lambda _{R1}}{\mu ^2}\right)-\frac{3}{2}\right)\,.
\end{split}
\end{equation}

The Goldstone contributions (9 d.o.f.) to the one-loop effective potential $V_{\text{Gold}}$ are:
\begin{equation}
\frac{9}{64 \pi ^2} \left(\rho ^2 \left(\lambda _{R1}+\lambda _{R2}\right)\right)^2 \left(\log \left(\frac{\rho ^2 \left( \lambda _{\text{R1}}+\lambda _{\text{R2}}\right) }{\mu ^2}\right)-\frac{3}{2}\right)
\end{equation}

The lepto-quark contributions from $SU(4)$ gauge fields (6 lepto-quark $\times3$ polarization=$18$ d.o.f.) to the one-loop effective potential are:
\begin{equation}
V_{\text{lepto}}=\frac{18}{64 \pi ^2} \left(\frac{1}{4} g_4^2 \rho ^2\right)^2 \left(\log \left(\frac{g_4^2 \rho ^2}{4 \mu ^2}\right)-\frac{5}{6}\right)\,,
\end{equation}
where the tree level lepto-quark mass is given by $M_{\rm{lepto}}^2=\frac{1}{4}g_4^2v^2$,
and $g_4~(g_R) $ is the $SU(4)~(SU(2)_R)$ gauge coupling.
The gauge boson $W_R^{\pm}$ contributions (2 $W_R$ $\times3$ polarization=$6$ d.o.f.) to the one loop effective potential are:
\begin{equation}
V_{W_R^{\pm}}=\frac{6}{64 \pi ^2} \left(\frac{1}{4} g_R^2 \rho ^2\right)^2 \left(\log \left(\frac{g_R^2 \rho ^2}{4 \mu ^2}\right)-\frac{5}{6}\right)\,,
\end{equation}
where the tree level $W_R$ mass is given by $M_{W_R^{\pm}}^2=\frac{1}{4}g_R^2v^2$.

The $Z'$ boson contribution (1 $Z'$ $\times3$ polarization=$3$ d.o.f.) to the one loop effective potential $V_{Z'}$ is:
\begin{equation}
\frac{3}{64 \pi ^2} \left(\frac{1}{8} \left(2g_R^2 +3g_4^2\right)\rho ^2\right)^2 \left(\log \left(\frac{\left(2g_R^2 +3g_4^2\right) \rho ^2}{8 \mu ^2}\right)-\frac{5}{6}\right)\,,
\end{equation}
where the tree level $Z'$ mass is given by $M_{Z'}^2=\frac{1}{8} \left(2g_R^2 +3g_4^2\right)v^2$.

The neutrino singlet contribution ($4$ d.o.f.~of Dirac Fermion) to the one loop effective potential is:
\begin{equation}
V_{\rm{\nu}}=-\frac{4}{64 \pi ^2} \left(\frac{1}{2}y_{\nu}^2\rho ^2\right)^2 \left(\log \left(\frac{y_{\nu}^2\rho ^2}{2 \mu ^2}\right)-\frac{3}{2}\right)\,,
\end{equation}
where the tree level neutrino singlet mass is given by $M_{\rm{\nu}}^2=\frac{1}{2}y_{\nu}^2v^2$

On the other hand, the Yukawa coupling in Eq.~\eqref{YukF} also contributes to the potential
as (4 colours $\times4$ d.o.f. of Dirac fermion=$16$ d.o.f.):
\begin{equation}
V_{\rm{F}}=-\frac{16}{64 \pi ^2} \left(\frac{1}{2}y_{F}^2\rho ^2\right)^2 \left(\log \left(\frac{y_{F}^2\rho ^2}{2 \mu ^2}\right)-\frac{3}{2}\right)\,,
\end{equation}
with a mass term $M_{F}^2=\frac{1}{2}y_{F}^2v^2$.
All in all, the total one-loop effective potential is:
\begin{equation}
V_{\rm{1loop}}=V_{\rm{Higgs}}+V_{\rm{Gold}}+V_{\rm{lepto}}+V_{W_R^{\pm}}+V_{Z'}+V_{\rm{\nu}}+V_F\,.
\end{equation}

\subsection{Finite Temperature Effective Potential of Pati-Salam Model}
The one loop finite temperature effective potential has the following general form
\begin{equation}
V_T=\sum_i\pm n_i \frac{T^4}{2\pi^2}\int_0^\infty dy y^2\log\left[1\mp e^{-\sqrt{y^2+m_i^2/T^2}}\right] \, ,\label{finite_temperature_full}
\end{equation}
where $+n_i~( - n_i)$ corresponds to bosons~(fermions).
We can further write the thermal integral in the form of the polynomials which can significantly simplify the calculations. We focus on the integral part of Eq.~\eqref{finite_temperature_full} and define:
\begin{equation}
I_{B,F}\left(a\right)=\pm\int_o^\infty dy y^2\log\left[1\mp e^{-\sqrt{y^2+a}}\right]\,,\label{thermal_integral}
\end{equation}
where we have used $a\equiv m_i^2/T^2$. For high temperature expansions ($m_i/T\ll1$), the thermal integral can be expanded respectively for bosons and fermions as:
\begin{equation}
\begin{split}
I_B^H\left(a\right)&=-\frac{\pi^4}{45}+\frac{\pi^2}{12}a-\frac{\pi}{6}a^{\frac{3}{2}}-\frac{a^2}{32}\left(\log\left(a\right)-c_B\right)\\
I_F^H\left(a\right)&=-\frac{7\pi^4}{360}+\frac{\pi^2}{24}a+\frac{a^2}{32}\left(\log\left(a\right)-c_F\right)\,,\label{high}
\end{split}
\end{equation}
where $c_B$ and $c_F$ are respectively $c_B=3/2-2\gamma_E+2\log\left(4\pi\right)$ and $c_F=3/2-2\gamma_E+2\log\left(\pi\right)$ and $\gamma_E\approx0.5772$. For low temperature expansions ($m_i/T\gg1$), the thermal integral for both bosons and fermions can be expanded as \footnote{Note that there are typos in the expressions of low energy expansion in \cite{Li:2014wia}.}:
\begin{equation}
I_{B,F}^L\left(a\right)=- \sqrt{\frac{\pi}{2}}a^{\frac{3}{4}}e^{- \sqrt{a}}
\left(1 +  \frac{15}{8} a^{- \frac{1}{2}} + \frac{105}{128} a^{-1}\right)\,.\label{low}
\end{equation}
To include the information for both high temperature and low temperature, we need to have an expression to connect the above two expressions Eq.~\eqref{high} and Eq.~\eqref{low}. We find:
\begin{equation}
\begin{split}
I_B\left(a\right)&=e^{-\left(\frac{a}{6.3}\right)^4}I_B^H\left(a\right)+\left(1-e^{-\left(\frac{a}{6.3}\right)^4}\right) I_{B}^L\\
I_F\left(a\right)&=e^{-\left(\frac{a}{3.25}\right)^4}I_F^H\left(a\right)+\left(1-e^{-\left(\frac{a}{3.25}\right)^4}\right) I_{F}^L\\
\end{split}
\end{equation}
Thus, we have the finite temperature effective potential (without ring contributions so far) as:
\begin{equation}
\begin{split}
V_T^{\rm{tot}}=&\frac{T^4}{2\pi^2}\left(I_B\left[\frac{M_{\rm{Higgs1}}^2}{T^2}\right]+6I_B\left[\frac{M_{\rm{Higgs2}}^2}{T^2}\right]\right.\\
+&9I_B\left[\frac{M_{\rm{Gold}}^2}{T^2}\right]+6I_B\left[\frac{M_{W_R^{\pm}}^2}{T^2}\right]+3I_B\left[\frac{M_{Z'}^2}{T^2}\right]\\
+&\left.18I_B\left[\frac{M_{\rm{lepto}}^2}{T^2}\right]+4I_F\left[\frac{M_\nu^2}{T^2}\right]+16I_F\left[\frac{M_{F}^2}{T^2}\right]\right)\,.
\end{split}
\end{equation}

\subsection{Ring Contribution to the Effective Potential of Pati-Salam Model}
The general formula for the ring contributions can be written as:
\begin{equation}
V_{\rm{ring}}^i=-\frac{T}{12\pi}\left(\left[m_i^2\left(\rho\right)+\sum_{\text{bosons}~j }
\pi_i^j\left(0\right)\right]^{3/2}-m_i^3\left(\rho\right)\right)\,,\label{V_ring}
\end{equation}
where $\pi_i\left(0\right)$ denotes the corresponding thermal mass contributions to the species $i$ from the relevant
bosonic d.o.f.~$j$ (in the outside rings of the daisy diagram). To consider the ring diagram contributions to the Higgs field, for example, $\pi_{\rm{Higgs}}$ should include all the scalar field (thermal mass) contributions denoted as $\pi_{\rm{Higgs}}^{\rm{Higgs1}},\,\pi_{\rm{Higgs}}^{\rm{Higgs2}},\,\pi_{\rm{Higgs}}^{\rm{Gold}}$
as well as the gauge field contributions.
For thermal mass contributions to the scalar field from the gauge and scalar fields (i.e.~scalar field in the big central ring of the Daisy diagram), we have the following general formula for the contributions of different species $j$ in the outside ring of the daisy diagram i.e.
\begin{equation}
\label{ring_general}
\pi_{\rm{scalar}}^j\left(0\right)=\frac{1}{12}\frac{m_j^2\left(v\right)}{v^2}T^2\,.
\end{equation} 
Thus, we obtain the thermal mass from the two Higgs fields and Goldstone fields respectively as:
\begin{equation}
\begin{split}
&\pi_{\rm{scalar}}^{\rm{Higgs1}}\left(0\right)=\frac{1}{4}\left(\lambda_{R1}+\lambda_{R2}\right)T^2,\quad \pi_{\rm{scalar}}^{\rm{Higgs2}}\left(0\right)=\frac{1}{12}\lambda_{R1}T^2\\
&\pi_{\rm{scalar}}^{\rm{Gold}}\left(0\right)=\frac{1}{12}\left(\lambda_{R1}+\lambda_{R2}\right)T^2\,.\label{scalar_scalar_Ring}
\end{split}
\end{equation}
Similarly, the scalar thermal mass contributions from the gauge fields are obtained in the following:
\begin{equation}
\begin{split}
&\pi_{\rm{scalar}}^{\rm{lepto}}\left(0\right)=\frac{1}{48}g_4^2T^2,\qquad \pi_{\rm{scalar}}^{W_R^{\pm}}\left(0\right)=\frac{1}{48}g_R^2T^2\\
&\pi_{\rm{scalar}}^{Z'}\left(0\right)=\frac{1}{96}\left(2g_R^2+3g_4^2\right)T^2\,.\label{gauge_scalar_Ring}
\end{split}
\end{equation}
To obtain the total thermal mass contributions to the Higgs field, we need to include all the above thermal masses i.e.~Eq.\eqref{scalar_scalar_Ring}, Eq.~\eqref{gauge_scalar_Ring} and 
we have:
\begin{equation}
\begin{split}
\sum_j\pi_{\rm{scalar}}^j\left(0\right)&=\pi_{\rm{scalar}}^{\rm{Higgs1}}\left(0\right)+6\pi_{\rm{scalar}}^{\rm{Higgs2}}\left(0\right)+9\pi_{\rm{scalar}}^{\rm{Gold}}\left(0\right)\\
&+18\pi_{\rm{scalar}}^{\rm{lepto}}\left(0\right)+6\pi_{\rm{scalar}}^{W_R^{\pm}}\left(0\right)+3\pi_{\rm{scalar}}^{Z'}\left(0\right)\,.
\label{scalar_TOT}
\end{split}
\end{equation}
Note that for each scalar field d.o.f.~(either the Higgs or Goldstone bosons), it receives the same ring diagram contributions $\sum_j \pi^j_i$.
Thus, by using Eq.~\eqref{V_ring} and Eq.~\eqref{scalar_TOT}, we obtain the total ring contributions to the scalar fields in the Pati-Salam model are:
\begin{equation}
V_{\rm{ring}}^{\rm{scalar,tot}}=V_{\rm{ring}}^{\rm{Higgs1}}+6V_{\rm{ring}}^{\rm{Higgs2}}+9V_{\rm{ring}}^{\rm{Gold}}\,.
\end{equation}

Now we consider the case where the gauge fields are in the central ring of the Daisy diagram. We have the following general formulas to calculate the gauge, scalar and fermion fields contributions to the gauge thermal masses for both abelian and Non-abelian cases. For abelian case, we have:
\begin{equation}
U\left(1\right):\quad \pi^{L,S}_{\rm{gauge}}=\frac{g'^2T^2}{3}\sum_S Y_S^2,\quad \pi^{L,F}_{\rm{gauge}}=\frac{g'^2T^2}{6}\sum_F Y_F^2\,,
\end{equation}
where $L$ denotes the longitudinal thermal mass since it can be shown that that the transverse thermal mass is suppressed and $Y_S,\,Y_F$ correspond respectively to the hypercharge of relevant scalar and fermion fields. For non-abelian case, we have:
\begin{equation}
\begin{split}
SU\left(N\right):\quad \pi^{L,S}_{\rm{gauge}}&=\frac{g^2T^2}{3}\sum_S t_2\left(R_S\right),\\
\pi^{L,F}_{\rm{gauge}}&=\frac{g^2T^2}{6}\sum_Ft_2\left(R_F\right),\\
\pi^{L,V}_{\rm{gauge}}&=\frac{N}{3}g^2T^2\,,
\end{split}
\end{equation}
where $t_2\left(R_S\right),\,t_2\left(R_F\right)$ corresponds respectively to the Dynkin indices of the scalar and fermion representations, $\text{Tr}[T_R^a T_R^b] =t_2(R) \delta^{ab}$.
We obtain the total thermal mass contributions to the lepto-quark, $W_R^{\pm}$, and $Z'$ are:
\begin{equation}
\pi^{L,\rm{Tot}}_{\rm{lepto}}=\frac{5}{3}g_4^2T^2,~~\pi^{L,\rm{Tot}}_{W_R^{\pm}}=\frac{4}{3}g_R^2T^2,~~\pi^{L,\rm{Tot}}_{Z'}=\frac{4}{3}g_4^2T^2\,.
\end{equation}
When computing ring contributions for gauge fields, we use the original basis instead of the mass eigenstates.
 Thus, both $m_i^2\left(\rho\right)$ and $\sum_i\pi_i^j\left(0\right)$ are rewritten as matrices $\mathbf{M^2\left(\rho\right)}$ and $\mathbf{\Pi\left(0\right)}$ respectively rather than eigenvalues as in the above scalar case. Eq.~\eqref{ring_general} can be correspondingly modified as:
\begin{equation}
V^{\text{gauge,tot}}_{\rm{ring}}=-\frac{T}{12\pi}\Tr\left(\left[\mathbf{M^2\left(\rho\right)}+\mathbf{\Pi\left(0\right)}\right]^{3/2}-\mathbf{M^3\left(\rho\right)}\right)\,.
\end{equation}
where we include all contributions to the gauge rings and take into account only the massive gauge bosons
for the big rings.
The $\mathbf{\Pi\left(0\right)}$ is a diagonal 10-by-10 matrix with the entries of $(i,i)$ being $5 g_4^2 T^2/3$ 
and entries of $(j,j)$ being $4 g_R^2 T^2/3$ for $i=(1, \dots,7)$ and $j = (8,9,10)$.

In contrast, $\mathbf{M^2\left(\rho\right)}$ is a nearly-diagonal symmetric 10-by-10 matrix with
the first six diagonal elements being $g^2_4 \rho^2/4$, the seventh being $3g^2_4 \rho^2/8$, and the last three diagonal
being $g^2_R \rho^2/4$, plus two off-diagonal elements: 
$ [\mathbf{M^2\left(\rho\right)} ]_{7,\,10}$$=$$[\mathbf{M^2\left(\rho\right)}]_{10,\, 7}$$= \sqrt{3/32} g_R g_4 \rho^2$.

\subsection{Complete Finite Temperature  Potential}

Now we are ready to write out the total finite temperature effective potential of the Pati-Salam model. It can be written as:
\begin{equation}
V_{\rm{tree}}+V_{\rm{1loop}}+V_T^{\rm{tot}}+V_{\rm{ring}}^{\rm{scalar,tot}}+V_{\rm{ring}}^{\rm{gauge,tot}}\,.\label{total_potential}
\end{equation}

\section{First Order Phase Transition and Gravitational Wave}
\label{PhaseTransition}

In this section, we will discuss the order of the possible early time Pati-Salam phase transition and the impact on possible gravitational wave signals.

\subsection{Strong First Order Phase Transition}
Here we focus on showing  that a strong first order phase transition can occur at around the Pati-Salam symmetry breaking scale with a sample coupling solutions shown in Tab.~\ref{sample solution_1}. 
\begin{table}[t!]
\centering
  \begin{tabular}{|| l | l | l | l | l | l | l | l | l | l | l ||}
    	\hline
  $\alpha_L$ & $\alpha_R$ & $\alpha_4$  &  $\lambda_{R1}$ & $\lambda_{R2}$ & $y_F$ & $y_\nu$ \\ \hline
   0.0038 & 0.0015 & 0.0109 & 0.291 & -0.291 & 0.004 & 0.645 \\ \hline
\end{tabular}
\caption{\small This table summarizes the sample coupling solutions at the Pati-Salam symmetry breaking scale. We did not include $\lambda_1,\,\lambda_2,\,\lambda_3,\,\lambda_4,\,$$\lambda_{{R\Phi_1}},\,$$\lambda_{{R\Phi_2}},$ $\,\lambda_{{R\Phi_3}},\,y,\,y_c$ since they are irrelevant in studying the finite temperature effective potential. Note that this set of solutions is obtained from a safe UV fixed point.}
\label{sample solution_1}
\end{table}
We did not include all the couplings in  the table since the remaining couplings are irrelevant in the analysis of our effective potential. We further note that the sample solutions in Tab.~\ref{sample solution_1}  are the ones leading to an asymptotically safe extension of the Pati-Salam model.   However, we will show that the occurrence of a first order phase transition is not limited to this set of
specific values of the couplings.

\begin{figure}[t!]
\centering
\includegraphics[width=0.8\columnwidth]{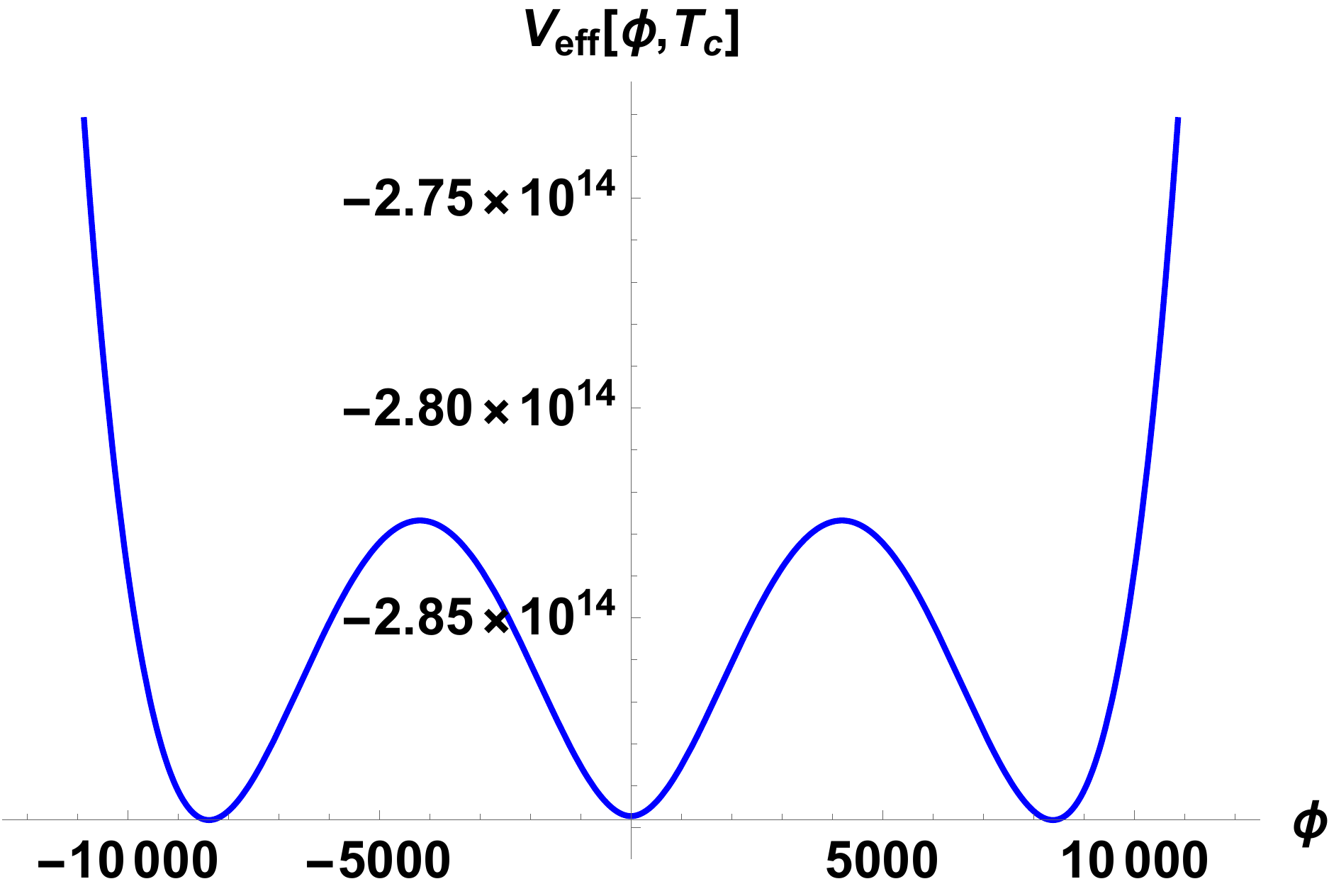}\hspace{0.07\columnwidth}
\caption{\small We plot the finite temperature effective potential by using the set of the couplings in Tab.~\ref{sample solution_1}. The renormalization scale $\mu$ is set at $5000\,\rm{TeV}$ while the temperature is chosen at $T=T_c=2680\,\rm{TeV}$ which is the critical temperature.}
\label{effective_potential}
\end{figure}

The finite temperature effective potential Eq.~\eqref{total_potential} is shown in Fig.~\ref{effective_potential}. Here we have set the renormalization scale $\mu$ at $5000\,\rm{TeV}$ that is  reasonable as the lower bound on the Pati-Salam physics scale is at  $2000\,\rm{TeV}$ or so, derived from the upper limit ${\rm Br}\left(K_L\rightarrow\mu^{\pm}e^{\mp}\right)<4.7\times10^{-12}$ \cite{Ambrose:1998us}. We have also chosen the temperature $T$ to match the critical temperature i.e.~$T=T_c=2680\,\rm{TeV}$ at which the potential has degenerate minima. 

A positive non-trivial (away from the origin) minimum  occurs for  $\phi_R\sim8400\,\rm{TeV}$ and it is denoted as $\phi_{Rc}$ and thus $\phi_{Rc}/T_c\sim 3.13>1$. This  shows  that the associated phase transition is a strong first order one.

\subsection{Connection between First Order Phase Transition and Coleman-Weinberg Symmetry Breaking}

We noticed that a strong first order phase transition occurs 
when spontaneous symmetry breaking happens via the Coleman-Weinberg mechanism. This is in line with the results and expectations of \cite{Sannino:2015wka}.  Of course, in other models first order phase transitions can still occur when symmetry breaking is generated via a hard negative mass square in the potential  \cite{Cline:1996mga}.

Around the finite temperature transition the Coleman-Weinberg values of the couplings reported in Tab.~\ref{sample solution_1} are such that  $\lambda_{R1} \simeq - \lambda_{R2}$ canceling each other. From the Renormalization Group (RG) flow point of view, Coleman-Weinberg symmetry breaking occurs when the RG flows of $\lambda_{R1}\left(\mu\right)+\lambda_{R2}\left(\mu\right)$ run from positive to negative flowing from the UV to the IR. The transition point (the scale $\lambda_{R1}\left(\mu\right)+\lambda_{R2}\left(\mu\right)=0$) defines the dynamical symmetry breaking scale of the Pati-Salam model which is  below $10000\,\rm{TeV}$.

\begin{figure}[t!]
\centering
\includegraphics[width=0.8\columnwidth]{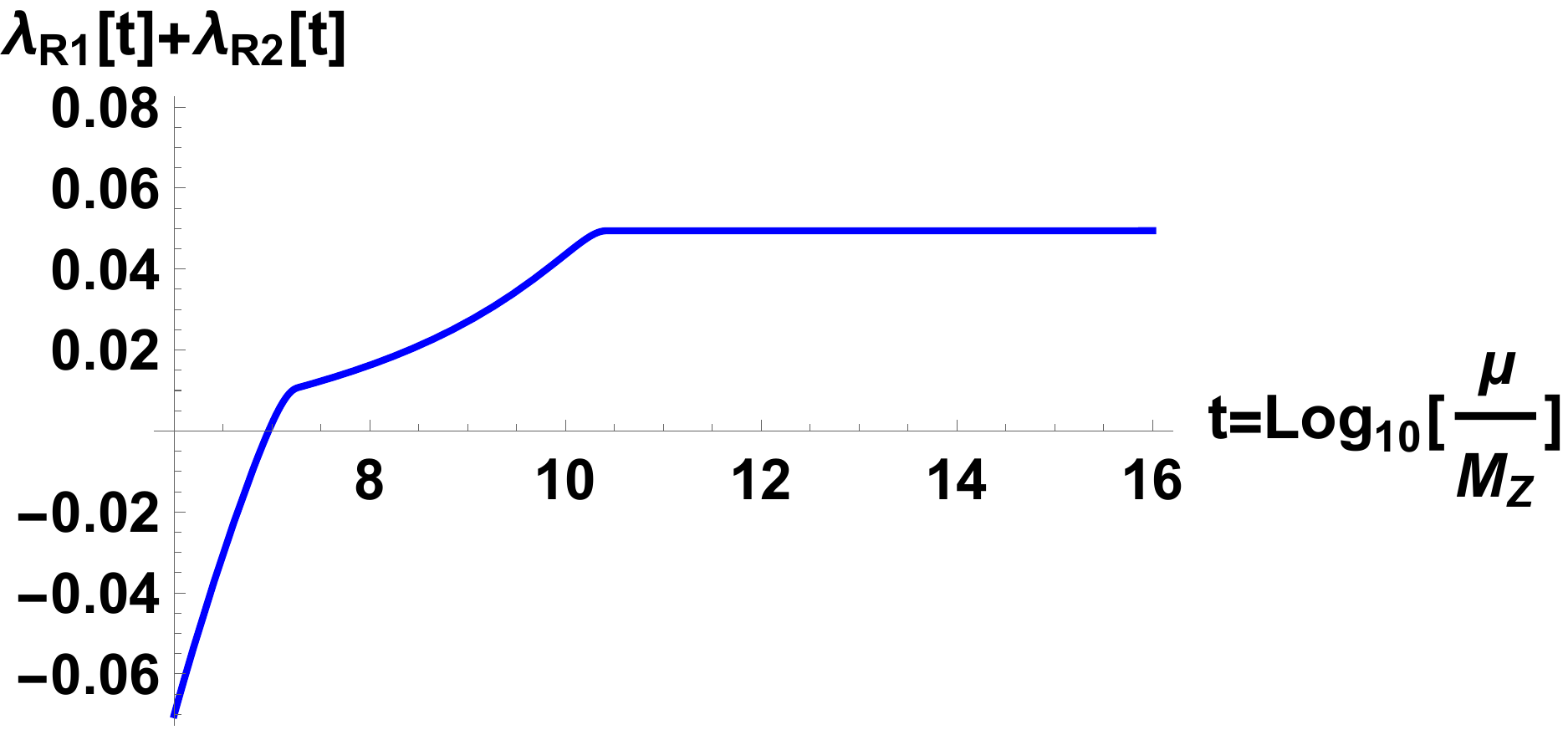}\hspace{0.07\columnwidth}
\caption{\small We plot the RG running of $\lambda_{R1}\left(t\right)+\lambda_{R2}\left(t\right)$ from UV to IR. The transition point (the scale $\lambda_{R1}\left(t\right)+\lambda_{R2}\left(t\right)=0$) defines the Coleman-Weinberg symmetry breaking scale of the Pati-Salam model.}
\label{Coleman_Weinberg}
\end{figure}

\begin{figure}[t!]
\centering
\includegraphics[width=0.8\columnwidth]{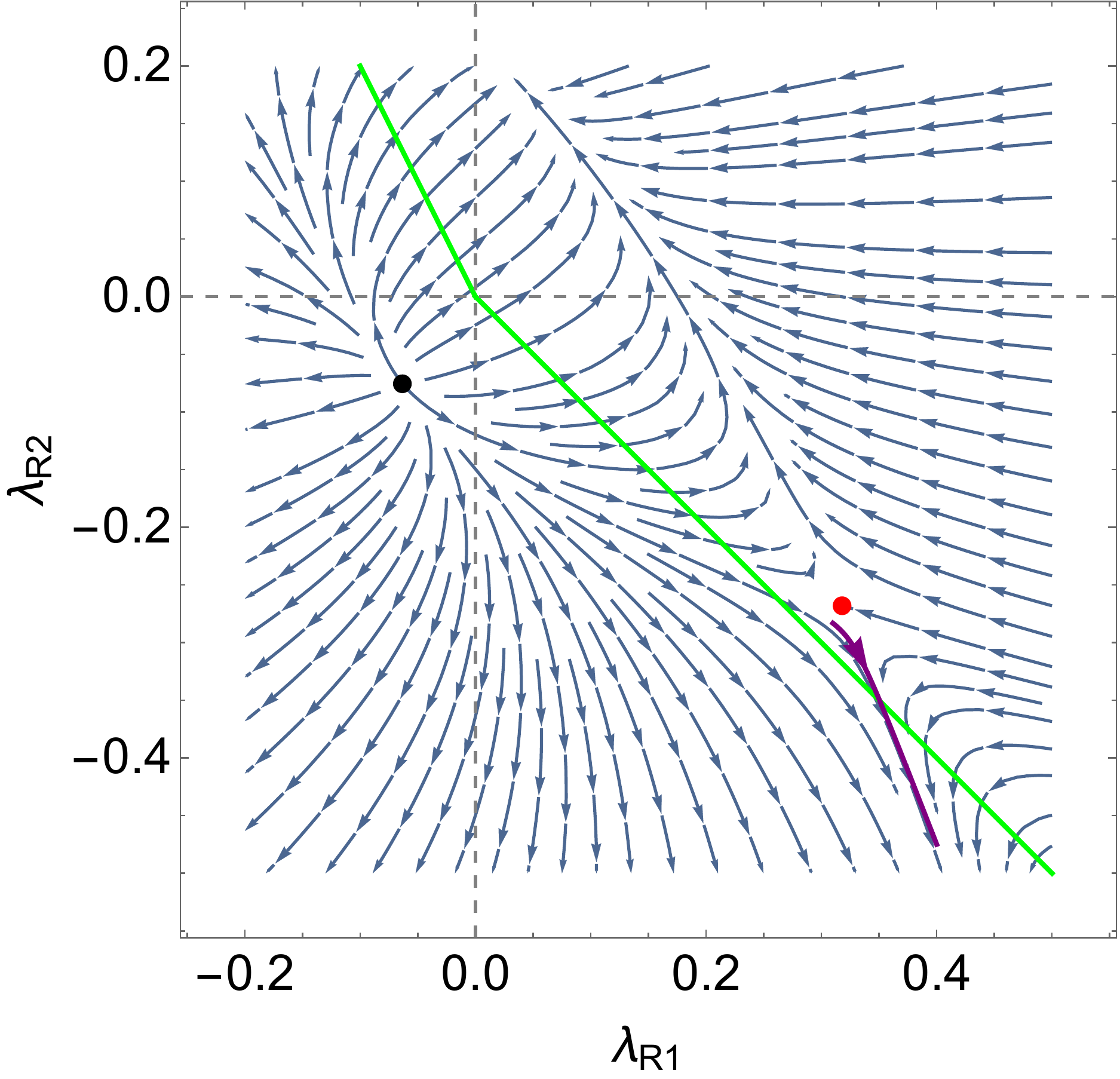}\hspace{0.07\columnwidth}
\caption{\small We show the stream plot of $\lambda_{R1},\,\lambda_{R2}$ where the flow direction is defined from UV to IR. The red and black plots are both the fixed point. The two green lines are the symmetry breaking lines which are defined as $\lambda_{R1}+\lambda_{R2}=0$ for $\lambda_{R2}<0$ and $\lambda_{R2}/2+\lambda_{R1}=0$ for $\lambda_{R2}>0$. The purple line is the particular RG flow corresponding to the sample solution in Tab.~\ref{sample solution_1}.}
\label{stream_plot}
\end{figure}

To gain insight it is interesting to show the symmetry breaking phenomenon via the stream plot provided in Fig.~\ref{stream_plot}.
The green line consisting of two symmetry breaking lines~($\lambda_{R1}+\lambda_{R2}=0$ for $\lambda_{R2}<0$ and $\lambda_{R2}/2+\lambda_{R1}=0$ for $\lambda_{R2}>0$) divides the plot into two phases. The right hand side of the green line corresponding to the vacuum stable phase while the left side is related to the symmetry breaking phase. In our convention the arrows point towards the infrared. The two dots correspond respectively to a saddle point (the red one)  and to an UV fixed point in both couplings. The bare couplings are meant to be fixed at some high energy scale on the right hand side of the plot. A glance at the plot shows that the only consistent way to radiatively cross the green line is by initiating the flow in the bottom right corner of the plot. One might be tempted to cross it from left to right by starting near the black dot. However this scenario would lead to an unstable potential at high energies and therefore is discarded. 

Focussing on the bottom right corner there is a special asymptotically safe trajectory emanating from the red dot.  On that trajectory the theory will avoid a Landau pole and can be considered fundamental (up to gravity) in the deep ultraviolet. Another point is that the trajectory leads to a predictive infrared physics. We are also pleased to see that there is a wider region of UV bare couplings values that lead to a Coleman-Weinberg phenomenon beyond the asymptotically safe limit.

\subsection{Bubble nucleation}

The time is ripe to discuss bubble nucleation within our model. We will provide a brief review of the method and apply it to our case. 

\begin{figure}[t!]
\centering
\includegraphics[width=0.8\columnwidth]{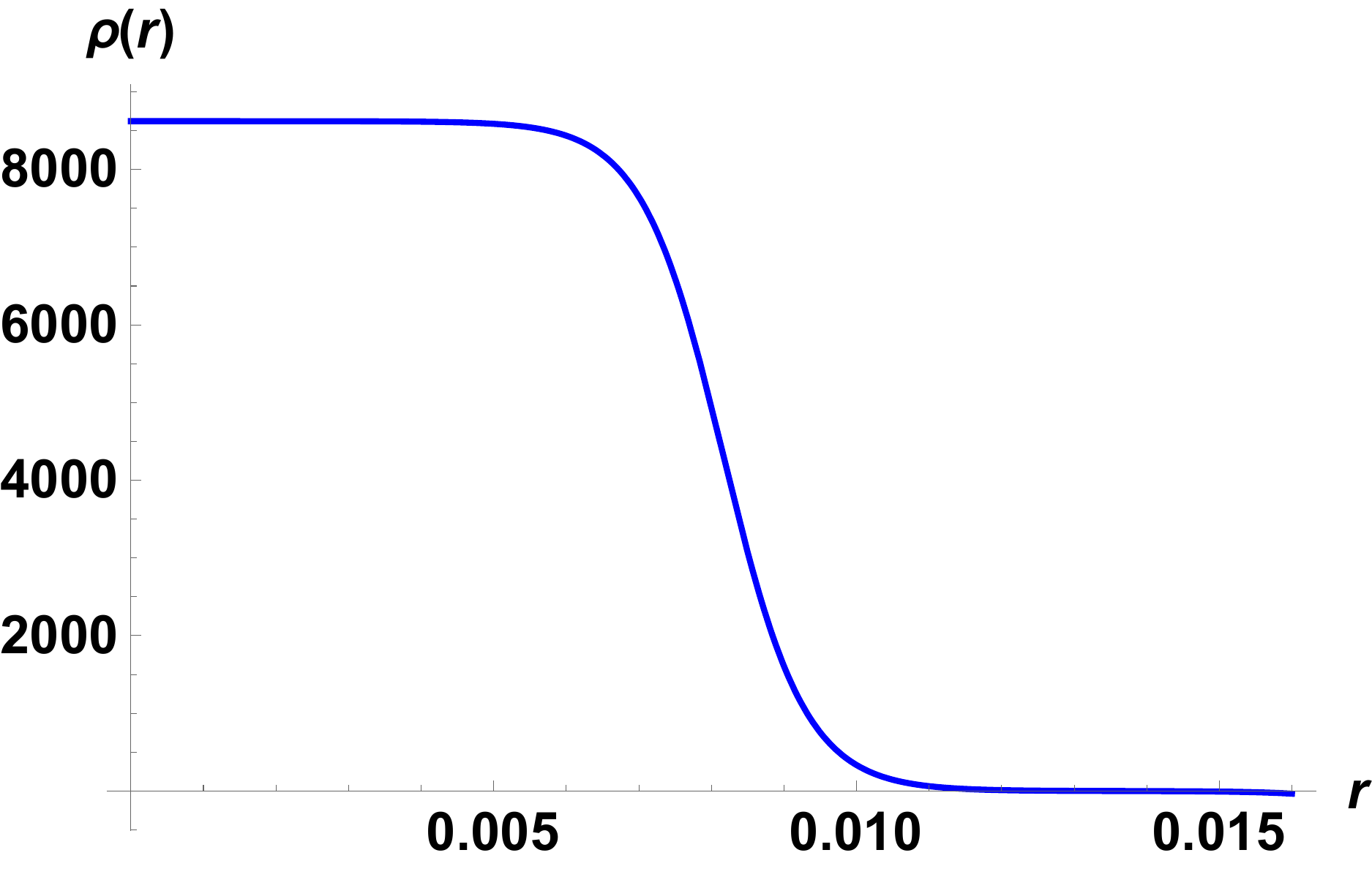}\hspace{0.07\columnwidth}
\caption{\small We plot the bubble profile $\rho\left(r\right)$, where $T$ is chosen at $T=2200\,\rm{TeV}$ which is slightly lower than the critical temperature at $T_c=2680\,\rm{TeV}$.}
\label{instanton_solution}
\end{figure}

The general picture is that as the universe cools down, a second minimum, away from the origin, develops below a critical temperature. This triggers the tunnelling from the false vacuum, at the origin,  to the stable vacuum below the critical temperature. Assuming the transition to be first order, the tunnelling rate per unit volume $\Gamma\left(T\right)$ from the metastable (false) vacuum to the stable one is suppressed by the three dimensional Euclidean action $S_3\left(T\right)$ and we have~\cite{Kobakhidze:2017mru}:
\begin{equation}
\Gamma\left(T\right)=\left(\frac{S_3\left(T\right)}{2\pi T}\right)^{3/2}T^4 e^{-S_3\left(T\right)/T}\label{decay_rate}
\end{equation}
The Euclidean action has the form:
\begin{multline}
S_3\left(\rho,T\right)=4\pi\int_0^\infty dr r^2\left[\frac{1}{2}\left(\frac{d\rho}{dr}\right)^2  +V\left(\rho,T\right) \right.
\\ 
\left. -V\left(0,T\right)\right]\,,
\label{Euclidean_Action}
\end{multline}
where we use the difference of the potential $F\left(\rho,T\right)\equiv V\left(\rho,T\right)-V\left(0,T\right)$ to adjust the ``datum point" of the potential at zero. The bubble configuration (instanton solution) is give by solving the following equation of motion of  the action in Eq.~\eqref{Euclidean_Action}:
\begin{equation}
\frac{d^2\rho}{dr^2}+\frac{2}{r}\frac{d\rho}{dr}-\frac{\partial F}{\partial\rho}\left(\rho,T\right)=0\,,
\end{equation}
with the associated boundary conditions:
\begin{equation}
\frac{d\rho}{dr}\left(0,T\right)=0,\qquad \lim_{r\rightarrow\infty} \rho\left(r,T\right)=0\,.
\end{equation}
To find the solutions we use the so called overshooting and under shooting method. We also used the numerical package, CosmoTransitions~\cite{Wainwright:2011kj} to cross-check our results.
 For $T=2200\,\rm{TeV}$ the bubble profile is shown in  Fig.~\ref{instanton_solution}.
We can insert the bubble profile $\rho\left(r,T\right)$ into the Euclidean action Eq.~\eqref{Euclidean_Action} and thus $S_3$ will be  dependent on $T$ only. 

The next step is to obtain the nucleation temperature which is defined as the temperature at which the rate of bubble nucleation per Hubble volume and time is approximately one. This means:
\begin{equation}
\Gamma\left(T\right)\sim H^4\,,
\end{equation}
where $H$ is the Hubble constant. By using Eq.\eqref{decay_rate}, we obtain:
\begin{equation}
T\ln\frac{T}{m_{pl}}\simeq -\frac{S_3\left(T\right)}{4}\,,\label{nucleation}
\end{equation}
where $m_{pl}$ is the Planck mass.
By solving Eq.~\eqref{nucleation} numerically, we find the nucleation temperature $T_n$ is around $1260\,\rm{TeV}$.
The inverse duration of the phase transition $\beta$ relative to the Hubble rate $H_{*}$ at the nucleation temperature $T_n$ is given by:
\begin{equation}
\frac{\beta}{H_*}=\left[T\frac{d}{dT}\left(\frac{S_3\left(T\right)}{T}\right)\right]\bigg\vert_{T=T_n}\,.\label{beta}
\end{equation}
We numerically obtain $\beta/H_* \simeq 183$.

Next, we will calculate another important parameter $\alpha$ which is the ratio of the latent heat released by the phase transition normalized against the radiation density:
\begin{equation}
\begin{split}
\alpha&=\frac{\epsilon}{\rho_{\rm{rad}}}=\frac{1}{\frac{\pi^2}{30}g_*T_n^4}\left(-\Delta V+T_n\Delta s\right)\\
\Delta V&=V\left(v_{T_n},T_n\right)-V\left(0,T_n\right)\\
\Delta s&=\frac{\partial V}{\partial T}\left(v_{T_n},T_n\right)-\frac{\partial V}{\partial T}\left(0,T_n\right)\,,\label{alpha}
\end{split}
\end{equation}
where $v_{T_n}$ is the vacuum expectation value of the finite temperature effective potential at the nucleation temperature, and $g_*$~(=150) is the relativistic d.o.f. in the universe. 
We find $\alpha_{T_n}\equiv
\alpha(T=T_n) =0.217$.

\subsection{Gravitational Waves}

We are now have all the instruments to address the generation and potential observation of gravitational waves stemming from the Pati-Salam early times phase transition. 

For the reader's benefit we provide a brief review of the ingredients needed to discuss the acoustic gravitational waves signals by following Ref.~\cite{Weir:2017wfa}. The  discussion about collision dynamics of scalar field shells and turbulence can be found in ~\cite{Weir:2017wfa} and their effects can be safely neglected in light  of being sub-leading. 

 The power spectrum of the acoustic gravitational wave is given by:
\begin{equation}
h^2\Omega_{sw}\left(f\right)=8.5*10^{-6}\left(\frac{100}{g_*}\right)^{\frac{1}{3}}\Gamma_{AI}^2 \overline{U}_f^4\left(\frac{H_*}{\beta}\right)v_w S_{sw}\left(f\right)\,,\label{power spectrum}
\end{equation}
where the adiabatic index $\Gamma_{AI}=\overline{\omega}/\overline{\epsilon}\simeq 4/3$. $\overline{\omega}$ and $\overline{\epsilon}$ denote respectively the volume-averaged enthalpy and energy density respectively. $\overline{U}_f$ is a measure of the root-mean-square (rms) fluid velocity and is given by:
\begin{equation}
\overline{U}_f^2\simeq\frac{3}{4}\kappa_f \alpha_{T_n}\,,
\end{equation}
where $\kappa_f$ is the efficiency parameter and it is well approximated by
\begin{equation}
\kappa_f\sim\frac{\alpha}{0.73+0.083\sqrt\alpha+\alpha}
\end{equation}
when $v_\omega~(\text{wall speed}) \rightarrow 1$. The spectral shape $S_{sw}\left(f\right)$ is given by:
\begin{equation}
S_{sw}\left(f\right)=\left(\frac{f}{f_{sw}}\right)^3\left(\frac{7}{4+3\left(f/f_{sw}\right)^2}\right)^{\frac{7}{2}}
\end{equation}
with peak frequency $f_{sw}$ approximated by:
\begin{equation}
f_{sw}=8.9\mu\text{Hz}\,\frac{1}{v_\omega}\left(\frac{\beta}{H_*}\right)\left(\frac{z_p}{10}\right)\left(\frac{T_n}{100\,\rm{GeV}}\right)\left(\frac{g_*}{100}\right)^{\frac{1}{6}}
\end{equation}
with $z_p$ a simulation-derived factor that is of order $10$, and following \cite{Hindmarsh:2017gnf} we take it to be $6.9$.

\begin{figure}[t!]
\centering
\includegraphics[width=0.9\columnwidth]{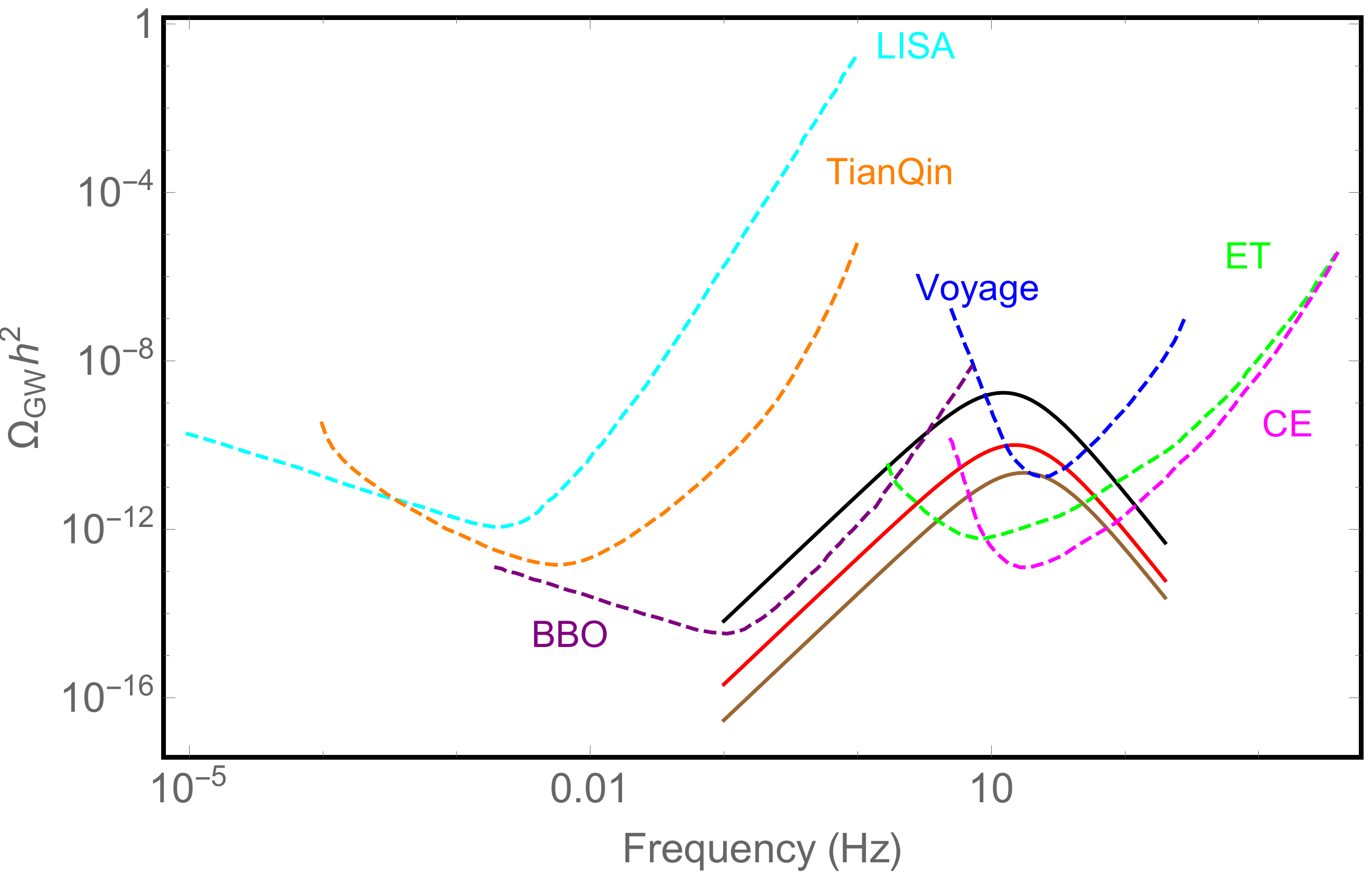}\hspace{0.09\columnwidth}
\caption{\small In this diagram, we show the bounds of the relevant future gravitational wave detections in the plot of dimensionless energy density in GWs against frequency. The bounds of LIGO Voyager, LISA, TianQing, BBO, Einstein Telescope (ET), Cosmic Explorer (CE) are shown respectively in blue, cyan, orange, purple, green and magenta.
The acoustic gravitational wave signals predicted in our Pati-Salam models are shown in black, red and brown. The red one is predicted in the asymptotically safe scenario while the black and brown ones are with the Yukawa coupling ($y_F, y_\nu$) values deviated from the safe scenario without modifying the IR SM physics. }
\label{voyager}
\end{figure}

By substituting $\alpha_{T_n}$ and $\beta/H_*$ from Eq.~\eqref{beta} and Eq.~\eqref{alpha} into the above power spectrum formula for the acoustic gravitational wave Eq.~\eqref{power spectrum}, we plot the curves of energy density against frequency (solid lines and the sample solution in Tab.~\ref{sample solution_1} is in red) in Fig.~\ref{voyager} where the coupling solutions in Tab.~\ref{parameter_space_1} are used. 
We have also included the future bounds~(dashed lines) coming from planned gravitational wave detection experiments such as LIGO Voyager \cite{Evans:2016mbw,Yagi:2017zhb}, LISA \cite{Caprini:2015zlo}, TianQing\footnote{The project's name consists of two Chinese words: "Tian", meaning sky or heavens, and "Qin", meaning the stringed instrument.}\cite{Luo:2015ght}, BBO \cite{Yagi:2011wg,Thrane:2013oya}, Einstein Telescope (ET) \cite{Punturo:2010zz,Hild:2010id} and Cosmic Explorer (CE) \cite{Evans:2016mbw}. They are shown respectively in blue, cyan, orange, purple, green and magenta in Fig.~\ref{voyager}. Interestingly, we find the predicted acoustic gravitational wave signal predicted  to be within the detection region of LIGO Voyager which is planned to be operational around 2027-2028.

\section{Pati-Salam driven Gravity Waves}
\label{GravityWaves}

We are now in a position to analyse in more detail the parameter space of bare couplings leading to observable gravitational waves within the Pati-Salam grand unified framework. 

For convenience we start with the asymptotically safe Pati-Salam scenario that has helped us quickly identify the relevant parameter space for the occurrence of a strong first order phase transition.

\subsection{Asymptotically Safe Case}
 In this section, we discuss an asymptotically safe embedding of the Pati-Salam framework by adding a large  number of vector-like fields into the theory.  In this limit we will argue for the existence of an UV fixed point  which solves the triviality problem while yielding a highly predictive theory at lower energies. 

Without further ado we introduce $N_F$ pairs of vector-like fermions charged under the fundamental representation of the Pati-Salam gauge group Eq.~\eqref{Pati_Group} with the following charge assignments:
\begin{equation}
N_F: \left(4,1,2\right)\oplus\left(4,2,1\right)\,.
\end{equation}
For simplicity, we assume that these new vector-like fermions appear at the Pati-Salam symmetry breaking scale.

\begin{table}[t!]
\centering
  \begin{tabular}{|| l | l | l | l | l | l | l | l | l | l | l | l ||}
    	\hline
   	$\lambda_1$ & $\lambda_2$ & $\lambda_3$ & $\lambda_4$ & $\lambda_{R\Phi_1}$ & $\lambda_{R\Phi_2}$ & $\lambda_{R1}$ & $\lambda_{R2}$ & $y,\,y_c$ & $y_F$ & $y_\nu$\\ \hline
   	0.13 & 0.01 & 0.03 & 0.05 & 0.10 & 0.01 & 0.34 & -0.29 & 0.53 & 0 & 0.67 \\ \hline
\end{tabular}
\caption{\small This table summarizes the UV fixed point solution for $N_F=13$ 
involving the bubble diagram contributions in the Yukawa and quartic RG beta functions. $y_{F}$ is asymptotically free and thus is zero at the fixed point.
}
\label{shifting UV fixed point_1}
\end{table}

Employing the large $N_F$ beta functions reported in appendix \ref{RG_beta} we can compute the RG flow connecting the UV fixed point (red dot in Fig.~\ref{stream_plot}) and the the SM in the infrared. For each $N_F\gg 1$ input, we obtain a set of UV fixed point solutions. Follow the RG flow starting from the determined UV fixed point to the electroweak scale, we can check whether it matches onto the SM.

At the PS symmetry breaking scale, we need to use matching conditions for both the gauge couplings and scalar quartic couplings. In particular, after PS symmetry breaking, the scalar bi-doublet should match the conventional two Higgs doublet model (we implement the beta functions of the two Higgs doublet model provided in \cite{Branco:2011iw}). We have searched the full parameter space in the range of $N_F\in\left(10,\,200\right)$ and find that $N_F=13$ with the UV fixed point solutions shown in Tab.~\ref{shifting UV fixed point_1} agree best with the low energy data (both the Higgs mass and the top Yukawa coupling at the electroweak scale).
We note that  $y_F$ is  asymptotically free for all viable solutions. We have therefore provided a UV safe completion of the SM \footnote{We note that even if the fixed point is not entirely established this analysis is still valid because the associated trajectories are valid for any energy scale sufficiently close to the would-be UV fixed point due to the nature of the precise results of the large $N_f$ expansion away from the fixed point.}.

 The sample solutions in Tab.~\ref{sample solution_1} are already the asymptotically safe solutions corresponding to $N_F=13$. This set is particularly interesting because:
\begin{itemize}
\item It corresponds to a possible UV safe fixed point rendering (up to gravity) our Pati-Salam model UV complete.
\item The Pati-Salam symmetry is dynamically broken through the Coleman-Weinberg mechanism below $10000\,\rm{TeV}$ (see Fig.~\ref{Coleman_Weinberg}) without adding any mass terms\footnote{This result does not depend on the existence of the fixed point but it is a welcome prediction.}.
\item Below $2680\,\rm{TeV}$
a strong first order phase transition occurs and at the nucleation temperature $T_n=1260\,\rm{TeV}$ gravitational wave signals can be generated. These are within the reach of the planned LIGO Voyager experiment detection region (see Fig.~\ref{voyager}) as well as the detection regions envisioned for the Einstein Telescope (ET), Cosmic Explorer (CE) and Big Bang Observer (BBO).
 \end{itemize}

We show the results as the red solid curve in both Fig.~\ref{voyager} and Fig.~\ref{bound_2}).

\subsection{Beyond the safe scenario}
Here, we will go beyond the safe scenario by exploring a more general parameter space able to generate testable gravitational wave signals.

We observe that the gauge couplings $g_4,\,g_R,\,g_L$ are  fixed by the Standard Model once the Pati-Salam symmetry breaking scale is chosen. In addition, when varying the quartic couplings we must ensure the presence of the Standard Model Higgs with its $125\,\rm{GeV}$  mass at the electroweak scale.
We therefore vary only the Yukawa couplings $y_F,\,y_\nu$ and the two quartic couplings $\lambda_{R1},\,\lambda_{R2}$ to satisfy this constraint. 

Scanning the Yukawa coupling parameter space, we discover that  when increasing either $y_F$ or $y_\nu$ (see black row of Tab.~\ref{parameter_space_1}), the dimensionless energy density of the gravitational wave signal  increases accordingly and the peak frequency will shift slightly to the left. This is clear when comparing the black curve with the red~(safe) curve in Fig.~\ref{voyager}. 

\begin{figure}[t!]
\centering
\includegraphics[width=0.9\columnwidth]{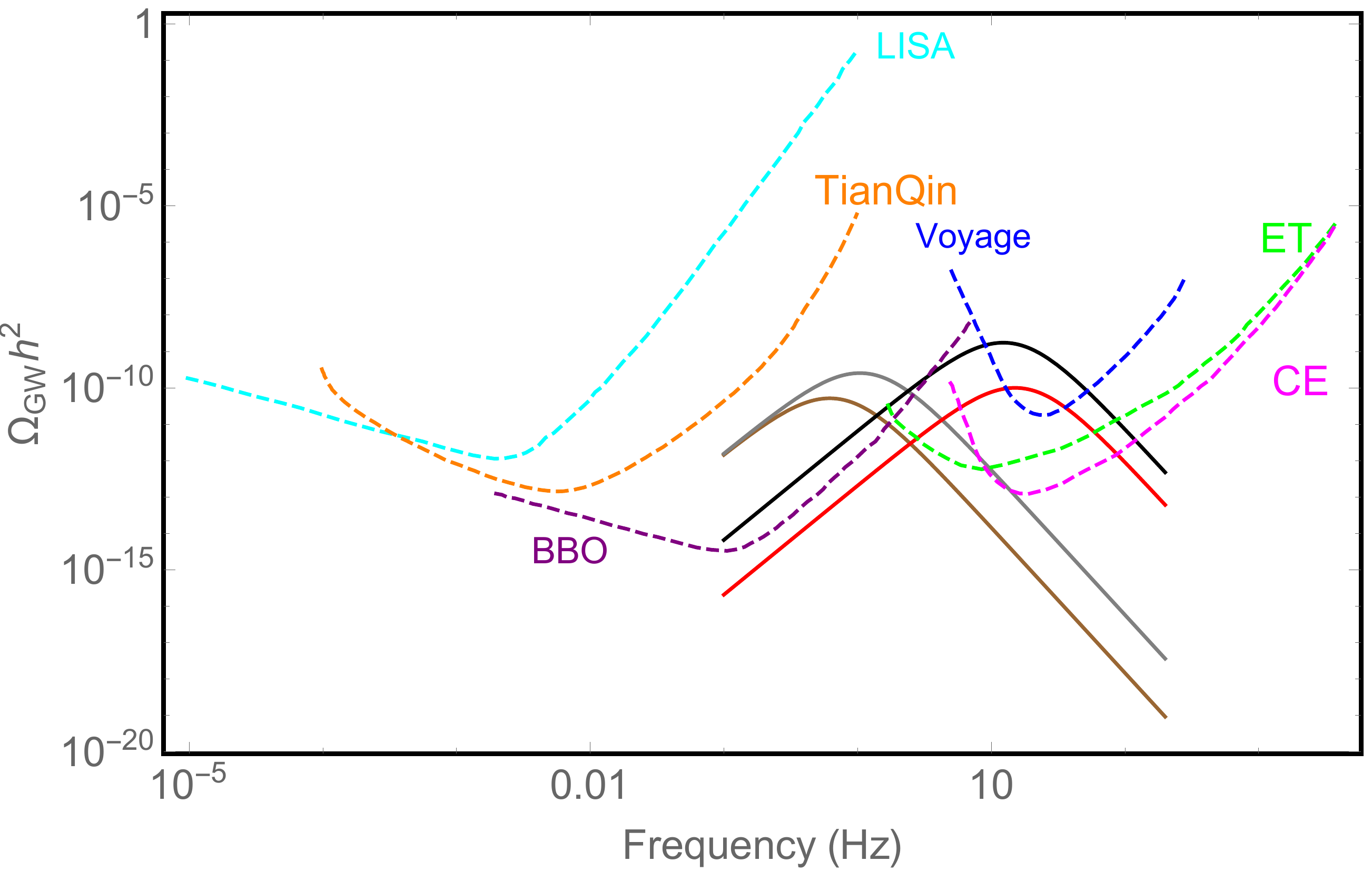}\hspace{0.09\columnwidth}
\caption{\small In this diagram, we show the bounds of the relevant future gravitational wave detections in the plot of dimensionless energy density in GWs against frequency. The bounds of LIGO Voyager, LISA, TianQing, BBO, Einstein Telescope (ET), Cosmic Explorer (CE) are shown respectively in blue, cyan, orange, purple, green and magenta.
The acoustic gravitational wave signals predicted in our Pati-Salam models are shown in red, black, grey and brown. The red one is predicted in the asymptotically safe scenario while the black, grey and brown ones are with the Yukawa and Quartic coupling ($\lambda_{R1}, \lambda_{R2}$) values deviated from the safe scenario without touching the IR SM physics. }
\label{bound_2}
\end{figure}

When scanning the quartic couplings parameter space, we find that the gravitational waves signal also depends on $\lambda_{R1}+\lambda_{R2}$. Varying $\lambda_{R1},\,\lambda_{R2}$ with fixed $\lambda_{R1}+\lambda_{R2}$, the dimensionless energy density and the peak of the frequency of the gravitational wave signals are roughly fixed. When increasing $\lambda_{R1}+\lambda_{R2}$ (see Brown and Grey row of Tab.~\ref{parameter_space_2}) the dimensionless energy density of the gravitational wave signal decreases accordingly and the peak frequency shifts significantly to the left with respect to the safe scenario. This can be seen  from Fig.~\ref{bound_2}.

Thus, differently from the safe scenario where the peak frequency is roughly  around $10$Hz,  going beyond the safe scenario allows for a peak of frequency ranging between $0.1$ and $10$Hz.

\begin{table}[t!]
\centering
  \begin{tabular}{|| l | l | l | l | l | l | l | l | l | l | l | l ||}
    	\hline
  & $\alpha_L$ & $\alpha_R$ & $\alpha_4$  &  $\lambda_{R1}$ & $\lambda_{R2}$ & $y_F$ & $y_\nu$ \\ \hline
   Safe &0.0038 & 0.0015 & 0.0109 & 0.291 & -0.291 & 0.004 & 0.645 \\ \hline
 Black &0.0038 & 0.0015 & 0.0109 & 0.291 & -0.291 & 0.5119 & 0.645 \\ \hline
 Brown &0.0038 & 0.0015 & 0.0109 & 0.291 & -0.291 & 0.001 & 0.001 \\ \hline
\end{tabular}
\caption{\small This table summarizes the sample coupling solutions at the Pati-Salam symmetry breaking scale. The Safe, Black, Brown represent respectively the gravitational wave energy density curves with Red, Black, Brown colours in Fig.~\ref{voyager}.}
\label{parameter_space_1}
\end{table}

\begin{table}[t!]
\centering
  \begin{tabular}{|| l | l | l | l | l | l | l | l | l | l | l | l ||}
    	\hline
  & $\alpha_L$ & $\alpha_R$ & $\alpha_4$  &  $\lambda_{R1}$ & $\lambda_{R2}$ & $y_F$ & $y_\nu$ \\ \hline
   Safe &0.0038 & 0.0015 & 0.0109 & 0.291 & -0.291 & 0.004 & 0.645 \\ \hline
 Black &0.0038 & 0.0015 & 0.0109 & 0.291 & -0.291 & 0.5119 & 0.645 \\ \hline
 Brown &0.0038 & 0.0015 & 0.0109 & 0.291 & -0.001 & 0.5119 & 0.645 \\ \hline
Grey &0.0038 & 0.0015 & 0.0109 & 0.291 & -0.093 & 0.5119 & 0.645 \\ \hline
\end{tabular}
\caption{\small This table summarizes the sample coupling solutions at the Pati-Salam symmetry breaking scale. The Safe, Black, Brown, Grey represent respectively the gravitational wave energy density curves with Red, Black, Brown, Grey colours in Fig.~\ref{bound_2}.}
\label{parameter_space_2}
\end{table}

\section{Conclusions}
\label{Conclusions}
We investigated the gravitational wave signatures stemming from the Pati-Salam model by identifying the parameter space of its couplings supporting a strong first order phase transition.

We started the analysis by employing a safe version of the Pati-Salam extension of the Standard Model and then quickly generalised to more generic situations. We find that a Coleman-Weinberg spontaneous breaking of the symmetry triggers a first order phase transition that can be observed via the next generation of gravitational wave detectors such as LIGO Voyager, the Einstein Telescope (ET) and the Cosmic Explorer (CE). 

Beyond the safe scenario we notice that the Yukawa couplings $y_\nu,\,y_F$ affect mostly the gravitational wave energy density while the combination of quartic couplings $\lambda_{R1}+\lambda_{R2}$ shifts its peak frequency. 

Concluding, we discover that the peak frequency of the gravitational wave signals stemming from the Pati-Salam model ranges within $0.1-10$ Hz.  Our results lead to the exciting news that the next generation of gravity waves detectors will be able to explore important extensions of the Standard Model appearing not at the electroweak scale but at much higher energy scales not accessible through present and future particle physics accelerators. 
   
\begin{acknowledgments}
The work is partially supported by the Danish National Research Foundation under the grant DNRF:90. 
WCH was supported by the Independent Research Fund Denmark, grant number DFF 6108-00623.
Zhi-Wei Wang thanks Tom Steele, Robert Mann and Steve Abel for helpful discussions and Huan Yang for recommending the relevant references.
\end{acknowledgments}
 

\appendix

\section{Large-N$_F$ beta functions}\label{RG_beta}
The beauty of large-N$_F$ beta function is by noticing that a subset of the Feynman diagrams (denoted as bubble chain) can be summed up into a closed form at $1/N_F$ order. Thus, all the higher order information up to $1/N_F$ order is encoded in the summation functions denoted as $F_1(A),\,H_1(A),\,H_0(A)$ below. It also deserves to note that these summation functions possess the pole structures:
\begin{equation}
F_1(A)\sim\log\left(1-2A/15\right),\quad H_1(A)\sim\log\left(1-A/3\right)\,,
\end{equation}
which guarantees the UV fixed point for the gauge beta functions and opens the possibility for the fixed point solutions for all the couplings.

To the leading $1/N_F$ order, the higher order (ho) contributions to the general RG functions of the gauge couplings  were computed in \cite{Antipin:2018zdg}, while for the simple gauge groups in \cite{Gracey:1996he,Holdom:2010qs} and for the abelian  in \cite{PalanquesMestre:1983zy}. Here we summarize the results. The ho contributions to $d \alpha_i/d\log\mu$ (in the semi-simple case) are:
\begin{equation}
\begin{split}
\beta^{{\rm ho}}_{ i}&=\frac{2A_i\alpha_i}{3}\left(\frac{d(R_i) H_{1_i}(A_i)}{N_{F_i} \,  \prod_{k} d\left(R_\psi^k\right) }+\frac{\sum_{j} \, d(G_j) \,F_{1_j}(A_j)}{N_{F_i} \prod_{k} d\left(R_\psi^k\right) }\right)\,,\\
\alpha_i&\equiv\frac{g_i^2}{\left(4\pi\right)^2}~~\left(i=L,\,R,\,C\right)\,,\label{higher order contribution}
\end{split}
\end{equation}
with the functions $H_{1i}$ and the t'Hooft couplings $A_i$  
\begin{equation}
\begin{split}
A_i&=4\alpha_iT_RN_{F_i}\frac{\prod_{k} d\left(R_\psi^k\right)}{d\left(R_\psi^i\right)}\\ 
H_{1_i}&=\frac{-11}{4}\frac{C_G}{T_R}+\int_0^{A_i/3}I_1(x)I_2(x)dx,\,\\
F_{1_j}&=\int_0^{A_j/3}I_1(x)dx ,\,\\ \label{summation_function_gauge}
\end{split}
\end{equation}
where $I_1(x)$ and $I_2(x)$ are:

\begin{equation}
\begin{split}
I_1(x)&=\frac{\left(1+x\right)\left(2x-1\right)^2\left(2x-3\right)^2\sin\left(\pi x\right)^3}{\left(x-2\right)\pi^3}\\
&\times\left(\Gamma\left(x-1\right)^2\Gamma\left(-2x\right)\right)\\
I_2(x)&=\frac{C_R}{T_R}+\frac{\left(20-43x+32x^2-14x^3+4x^4\right)}{4\left(2x-1\right)\left(2x-3\right)\left(1-x^2\right)}\frac{C_G}{T_R}\,.
\end{split}
\end{equation}
The Dynkin indices are $T_R=1/2~(N_{c_i})$ for the fundamental (adjoint) representation while $d\left(R_\psi^k\right)$ denotes the dimension of the fermion representation.

The RG functions of the (semi-simple) gauge couplings are:
\begin{equation}
\begin{split}
\beta_{\alpha_{2L}}^{tot}&=\frac{d\alpha_{2L}}{d\log\mu}=\beta_{\alpha_{2L}}^{1loop}+\beta_{\alpha_{2L}}^{\rm{ho}} =-6\alpha_{2L}^2 \\
&+\frac{2A_{2L}\alpha_{2L}}{3}\left(1 + \frac{H_{1_{2L}}\left(A_{2L}\right)}{4\, N_{F}}+\frac{15}{8} \frac{F_{1_4}\left(A_4\right)}{N_F}\right)\\
\beta_{\alpha_{2R}}^{tot}&=\frac{d\alpha_{2R}}{d\log\mu}=\beta_{\alpha_{2R}}^{1loop}+\beta_{\alpha_{2R}}^{\rm{ho}} =-\frac{14}{3}\alpha_{2R}^2 \\
&+\frac{2A_{2R}\alpha_{2R}}{3}\left( 1 + \frac{H_{1_{2R}}\left(A_{2R}\right)}{4\, N_{F}}+\frac{15}{8} \frac{F_{1_4}\left(A_4\right)}{N_F}\right)\\
\beta_{\alpha_{4}}^{tot}&=\frac{d\alpha_{4}}{d\log\mu}=\beta_{\alpha_{4}}^{1loop}+\beta_{\alpha_{4}}^{\rm{ho}} =-18\alpha_{4}^2 \\
+ & \frac{2A_{4}\alpha_{4}}{3}\left( 1 + \frac{H_{1_4}\left(A_{4}\right)}{4\, N_{F}}+ \sum_{i = L/R} \frac{3}{16} \left( \frac{F_{1_{2i}}\left(A_{2i}\right)}{N_F} \right)\right)\,,\\
\end{split}
\label{Gauge couplings RG Bubble}
\end{equation}
The Yukawa beta function reads
\begin{equation}
\begin{split}
\beta_y&=c_1y^3 + y \sum_{\alpha}c_\alpha g^2_\alpha I_y\left(A_\alpha\right),\quad\rm{with}\\\label{eq-simplifiedyukawa}
I_y\left(A_\alpha\right)&=H_\phi\left(0,\tfrac{2}{3}A_\alpha\right)\left(1+A_\alpha\frac{C_2\left(R_\phi^\alpha\right)}{6\left(C_2\left(R_{\chi}^\alpha\right)+C_2\left(R_{\xi}^\alpha\right)\right)}\right)\\
H_\phi(x) &=H_0(x)= \dfrac{(1 - \tfrac{x}{3}) \Gamma(4-x)}{3 \Gamma^2(2 - \tfrac{x}{2}) \Gamma(3 - \tfrac{x}{2}) \Gamma(1 + \tfrac{x}{2})}
\end{split}
\end{equation}
containing information about the resumed fermion bubbles and $c_1,\,c_\alpha$ are the standard 1-loop coefficients for the Yukawa beta function while $C_2(R_\phi^\alpha),\,C_2(R_{\chi}^\alpha),\,C_2(R_{\xi}^\alpha)$ are the Casimir operators of the corresponding scalar and fermion fields. Thus, when $c_1,\,c_\alpha$ are known, the full Yukawa beta function follows.
Similarly, for the quartic coupling we write
\begin{equation}
\begin{split}
\beta_\lambda&=c_1\lambda^2+\lambda \sum_{\alpha}c_\alpha \,g^2_\alpha\,I_{\lambda g^2}\left(A_\alpha\right)+\sum_{\alpha} c'_\alpha \,g_\alpha^4\,I_{g^4} \left(A_\alpha\right) \\
&+\sum_{\alpha < \beta}c_{\alpha\beta}\, g_\alpha^2 g_\beta^2 \,I^{tot}_{g_1^2g_2^2}\left(A_\alpha,\,A_\beta\right)\,,\label{quartic_bubble_beta}
\end{split}
\end{equation}
with $c_1,\,c_\alpha,\,c'_\alpha,\,c_{\alpha\beta}$  the known 1-loop coefficients
for the quartic beta function and the resumed fermion bubbles appear via 
\begin{equation}
\begin{split}
I_{\lambda g^2}\left(A_\alpha\right) &=H_\phi\left(0,\tfrac{2}{3}A_\alpha\right)\\
I_{g^4}\left(A_\alpha\right)&=H_\lambda\left(1,\tfrac{2}{3}A_\alpha\right)+A_\alpha\frac{dH_\lambda\left(1,\tfrac{2}{3}A_\alpha\right)}{dA_\alpha}
\end{split}
\end{equation}
\begin{equation}
\begin{split}
I_{g_1^2g_2^2}^{tot}\left(A_\alpha,\,A_\beta\right)&=\frac{1}{3}\bigg[I_{g_1^2g_2^2}\left(A_\alpha,\,0\right)+I_{g_1^2g_2^2}\left(0,\,A_\beta\right)\\
&+I_{g_1^2g_2^2}\left(A_\alpha,\,A_\beta\right)\bigg]\\
I_{g_1^2g_2^2}\left(A_\alpha,\,A_\beta\right)&=\frac{1}{A_\alpha-A_\beta}\bigg[A_\alpha H_\lambda\left(1,\tfrac{2}{3}A_\alpha\right)\\
&-A_\beta H_\lambda\left(1,\tfrac{2}{3}A_\beta\right)\bigg],\\
H_\lambda(1,x) &= (1-\tfrac{x}{4}) H_0(x)\\
&=\dfrac{ (1-\tfrac{x}{4})(1 - \tfrac{x}{3}) \Gamma(4-x)}{3 \Gamma^2(2 - \tfrac{x}{2}) \Gamma(3 - \tfrac{x}{2}) \Gamma(1 + \tfrac{x}{2})}\,.
\end{split}
\end{equation} 
We therefore have the  quartic beta function including the bubble diagram contributions when $c_1,\,c_\alpha,\,c_\alpha',\,c_{\alpha\beta}$ are known.

\end{document}